\documentclass{article}
\usepackage[utf8]{inputenc}
\usepackage{graphicx}
\graphicspath{{./images/}}

\usepackage{amssymb} 
\usepackage{amsmath}

\usepackage[square, numbers, sort&compress]{natbib}

\title{Decision Market Based Learning For Multi-agent Contextual Bandit Problems}
\author{Wenlong Wang and Thomas Pfeiffer}
\date{November 2022}

\begin{document}

\maketitle

\begin{abstract}
    Information is often stored in a distributed and proprietary form, and agents who own information are often self-interested and require incentives to reveal their information. Suitable mechanisms are required to elicit and aggregate such distributed information for decision making. In this paper, we use simulations to investigate the use of decision markets as mechanisms in a multi-agent learning system to aggregate distributed information for decision-making in a contextual bandit problem. The system utilises strictly proper decision scoring rules to assess the accuracy of probabilistic reports from agents, which allows agents to learn to solve the contextual bandit problem jointly. Our simulations show that our multi-agent system with distributed information can be trained as efficiently as a centralised counterpart with a single agent that receives all information. Moreover, we use our system to investigate scenarios with deterministic decision scoring rules which are not incentive compatible. We observe the emergence of more complex dynamics with manipulative behaviour, which agrees with existing theoretical analyses. 
\end{abstract}

\section{Introduction}
In many decision making tasks, the relevant information is distributed over multiple parties. To optimise decision making, multi-agent learning systems are required to obtain, aggregate and learn from such distributed information. When the agents’ information is private and the objective is self-interested, rewards may be required to induce the agents to reveal their information. For efficient multi-agent learning in such a situation, the rewards must be designed so that as agents maximise their rewards in the training phase, the system's overall performance is also optimised. 

Consider, for example, a recommendation system that aims to optimise advertisement targeting by using information from multiple sources (e.g., Google, Facebook and Amazon). Such information could involve the companies’ different user profile data for the targeted person, which the companies have no interest to reveal. The system, therefore, needs to elicit information in a form that is agreeable to the information source (e.g. recommendations for the task at hand, rather than complete user profiles) and needs to provide fair rewards for these contributions. Such rewards can be monetary but need to be designed such that each information source can learn from the realised rewards and while maximising its rewards, the performance of the recommendation system improves as well.

In this work, we develop a multi-agent learning system that provides  agents with rewards that align the agents’ objectives with the system’s objectives. We test the system in simulations of learning in a multi-armed Bandit problem where contextual information is distributed over multiple agents. Our approach is based on decision markets, which are an extension of prediction markets. While prediction markets are mechanisms of multi-agent forecasting, decision markets are mechanisms of multi-agent decision-making where decisions are made based on forecasts. The contextual bandit problem we study in the simulations can be seen as a one-step reinforcement learning problem.

This paper is organised as follows. In Section \ref{sec: related work}, we discuss relevant work from three related topics: multi-agent learning (Section \ref{sec: multi-agent learning}), bandit problems (Section \ref{sec: bandit problem}) and decision markets (Section \ref{sec: decision markets}). In Section \ref{sec: algorithm}, we introduce our research methodology. In Section \ref{sec: simulation results}, we present simulation results, and in Section \ref{sec: conclusion}, we discuss future work directions.

Our results show that the decision market based multi-agent bandit system can optimise decision making without requiring individual agents to share their contextual information directly. We use our system to examine the agents’ behaviour in a decision market with a stochastic decision rule which provides proper incentives for accurate reporting by the agents, and in a decision market with a deterministic decision rule which can be manipulated and exploited by the agents. We find that under a stochastic decision rule, agents quickly learn to  provide accurate reports. Under the deterministic decision rule, we observe interesting manipulative interactions that nevertheless often result in surprisingly good decision making.

\section{Related Work} \label{sec: related work}
\subsection{Multi-agent Learning} \label{sec: multi-agent learning}
Multi-agent learning is an essential and rapidly growing research area in computer science. According to the nature of the interactions between the agents, multi-agent learning algorithms can be grouped into three categories \cite{Busoniu2008ALearning}: purely cooperative, purely competitive, and mixed. 

An example of a cooperative task is the wolf-pack game where two wolves chase prey. Because the prey is faster than the wolfs, the wolves need to learn a cooperative strategy to capture the prey \cite{Lowe2017Multi-AgentEnvironments}. When either one wolf reaches the prey, the task is solved and the entire wolf-pack receives a reward. In such a cooperative task, agents usually share an identical reward function and thus learn to maximise the joint rewards of the system.

In competitive tasks, one agent’s gain is the other agent’s loss. In the wolf-pack game, for instance, the wolf agents and the prey agent are in a competitive game. This kind of game is well-studied in game theory. Optimal strategies can be learned with the min-max learning paradigm, which is guaranteed to find the best policy under the worst-case assumption of the opponent's move when the policy searching space is manageable \cite{Littman1994MarkovLearning}. Recent attention has focused on deep reinforcement learning to solve complex competitive games such as Go \cite{Silver2014}.

Mixed tasks contain elements of the two extremes described above and the nature of games and agents can be very diverse. For instance, some tasks involve two teams playing against each other \cite{Baker2020EmergentAutocurricula}, and cooperation and competition co-exist in these tasks. Some studies focus on the dynamics of self-interested agents’ interaction in such tasks \cite{Leibo2017Multi-agentDilemmas}. In our work, the agents engage in a partial information game that is neither cooperative nor  purely competitive. It is not cooperative, because the agents have individual reward functions and thus can be seen as self-interested, and they are not purely competitive because the overall reward provided to the agents increases if the agents work well together. To draw an analogy with the previous wolf-pack game, our game is essentially a wolf-crow game where a murder of crows has information about the location of numerous prey that the wolf would benefit from obtaining. The wolf needs to reward the crows for their information, such that they learn to guide the wolf to the most valuable prey.

A further classification of multi-agent learning exists along the training methods dimension. A significant share of attention focuses on a paradigm called centralised training decentralised execution \cite{Lowe2017Multi-AgentEnvironments, Rashid2018QMIX:Learning}. On the other hand, some algorithms fall into the fully decentralised paradigm. 

Centralised training with decentralised execution is suitable for agents that execute actions locally, based on local information. During the training phase, however, all the local information is accessible to a centralised ‘critic’ that can evaluate the agent’s actions from a higher level. This paradigm can ameliorate two problems in multi-agent learning: the ‘coordination problem’ and the ‘non-stationary issue’ \cite{Busoniu2008ALearning}. The coordination problem arises when agents have to ‘match’ or coordinate their actions to maximise rewards. The non-stationary issue arises when agents optimise in an environment that is non-stationary because it contains other co-optimising agents \cite{Lowe2017Multi-AgentEnvironments, Rashid2018QMIX:Learning}. This paradigm, however, is not suitable for our task, because the local information is private and is not readily shared.

Federated learning is a novel paradigm that aims to solve a centralised learning problem without compromising the privacy of users. The federated bandit problem \cite{Li2020FederatedPrivacy, Qi2021FederatedChallenges} provides an important framework for recommendation systems without central access to local data \cite{Shi2021FederatedBandits} and sometimes even without a centralised model \cite{Zhu2021FederatedApproach}. 

Another approach is to let agents learn local policies independently \cite{Tan1997Multi-AgentAgents, Tampuu2017MultiagentLearning}. This approach is suitable when information is private, but learning can be affected by the non-stationary problem. In our design, we use a mechanism from economics to decorrelate the relationship between the reward for an independent agent and the peers’ actions and therefore expect to mitigate the non-stationary problem.

\subsection{Bandit Problems} \label{sec: bandit problem}
Bandit problems provide a framework for studying optimal decision-making when several alternative actions are available that yield rewards from an unknown, stationary distribution. Actions can be discrete or continuous \cite{Agrawal1995TheProblem}. The rewards essentially quantify the quality of the selected action. Agents act repeatedly and learn by taking a history of the rewards into account for decision making in subsequent rounds. Agents need to balance the exploration for potential better action and the exploitation according to the best knowledge so far \cite{Lattimore2020BanditAlgorithms}. For some problems, ‘hints’ or contexts exist and provide information about the reward distribution associated with an action. The reward distribution is non-stationary and hints change accordingly. The bandit problem, therefore, extends to a contextual bandit problem which is equivalent to a one-step reinforcement learning problem \cite{Barto1985Pattern-recognizingAutomata}.

Contextual bandit problems are well-suited models for decision-making challenges based on current and past information. Many practical applications, such as product recommendations and advertisement targeting, are based on bandit learning and categorised into recommendation systems. An important use case of multi-agent bandit learning is cognitive radio. In cognitive radio, users want to identify idle channels intelligently, which is often discussed along with a multi-agent environment where two users want to avoid selecting the same idle channel \cite{Liu2010DistributedPlayers,Martinez-Rubio2019DecentralizedBandits,Bistritz2020MyBandits}. Recommendation systems are predominately investigated with centralised models, but with increasing regulation of data privacy, security and access right, decentralised models that can utilise the private data on individual devices are gaining relevance. In the system we are investigating, the agents learn locally from local contextual data to help optimise decision making in a multi-agent multi-armed bandit problem. 

\subsection{Decision Markets} \label{sec: decision markets}
Collective decision making with distributed information is a familiar challenge in economics. In decision markets, this challenge is addressed by eliciting and assessing forecasts about the consequences of the available actions. Specifically, a principal (decision maker) elicits forecasts from agents with access to relevant information and then selects an action according to these forecasts. After execution, the principal will compute scores for the agents’ contributions.

Scoring rules provide such an assessment of forecasts by assigning a real number score to forecasts depending on the realised outcome \cite{Gneiting2007StrictlyEstimation}. A proper score can guarantee the highest expected return if the evaluated forecast aligns with the actual belief of the forecaster. In other words, a rational agent maximising its expected score will under a proper scoring rule report the most accurate forecast it can make. Proper scoring rules are suited to reward single agents for their forecasts and allow them to learn to make forecasts more accurate.

While proper scoring rules provide proper incentives for single agents to make accurate forecasts, properly incentivised prediction markets are mechanisms to elicit and aggregate forecasts from multiple agents that have access to different pieces of information. The mechanisms of aggregation depend on the implementation of prediction markets. Hanson proposes a proper prediction market mechanism that allows agents to make direct probabilistic reports in a sequential order \cite{Hanson2003CombinatorialDesign}, and suggests market scoring rules to price assets such that sequential reporting and trading in an asset market becomes equivalent. Such a mechanism requires agents to make Bayesian updates based on the report from the previous agent. \citeauthor{Chen2007AMakers} further generalise market scoring rules and their relation to scoring rules \cite{Chen2007AMakers}. Implementations following this approach require a principal to provide liquidity for a prediction market by an automated market maker algorithm that is always available to trade.

While scoring rules assess single forecasts, and prediction markets provide a mechanism for aggregating the forecasts from multiple agents, decision markets extend these approaches to evaluate and aggregate forecasts for decision making \cite{Hanson1999DecisionMarkets}. The challenge is to quantify the quality of forecasts that are conditional on the actions. Because the realised future depends on the selected action, it is difficult to assess forecasts about the other actions. Decision rules, which choose actions according to multiple forecasts, are the core of the solution to this challenge. A naive way is to use a deterministic decision rule that always selects the best action according to the forecasts. However, this approach may incentivise misleading forecasts since the forecasts directly determine the selected action \cite{Othman2010DecisionMarkets, Chen2011InformationMaking}. \citeauthor{Chen2011DecisionIncentives} propose a stochastic decision rule that breaks this relation, and thus makes the forecasts reliable \cite{Chen2011DecisionIncentives, Chen2014ElicitingMaking}. \citeauthor{Wang2022SecuritiesMarkets} extend the mechanism from direct probabilistic forecasts to equivalent asset trading markets \cite{Wang2022SecuritiesMarkets}.

\section{Algorithm} \label{sec: algorithm}
\subsection{Problem Setup}
\begin{figure}
    \centering
    \includegraphics[width=\textwidth]{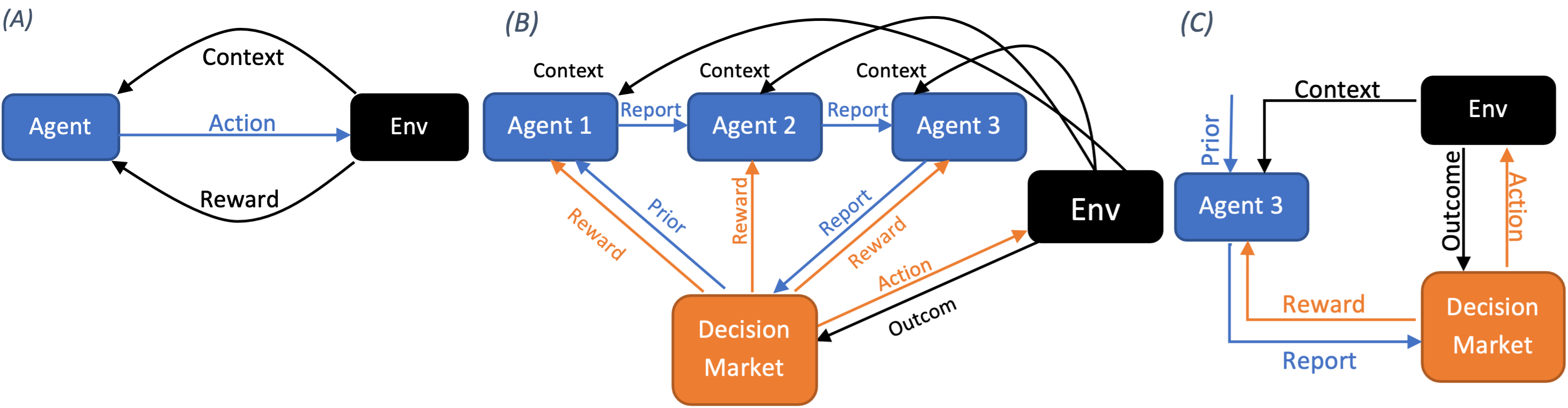}
    \caption{Decision markets based multi-agent bandit system. Panel (A) shows a diagram for a regular contextual bandit problem. An agent can choose an action after receiving contextual information. The action results in a reward from the environment. Panel (B) shows a multi-agent contextual bandit problem with a decision market, which is the main design of this paper. An action is selected by a decision market, which aggregates distributed posterior probabilities reported from agents. The decision market assigns a reward to each agent based on the quality of their reports. Panel (C) shows a contextual bandit problem with a continuum-arm space in agent 3’s perspective.}
    \label{fig:three_diagrams}
\end{figure}

We study a multi-agent multi-armed contextual Bernoulli bandit problem, where one agent (referred to as the principal) decides between multiple alternative actions and receives a corresponding reward that evaluates the quality of the decision. The context, however, is distributed over multiple self-interested agents. In the system we investigate here, the principal uses a decision market to sequentially elicit probabilistic reports for the Bernoulli outcomes of the available actions from the agents (see Figure \ref{fig:three_diagrams}). In each time step, the principal receives an initial set of prior probability distributions for the outcomes of each action. It then selects an agent to alter this report. The agent will be scored for this altered report using a decision scoring rule. The principal then adopts this report and selects the next agent to alter it, and this process is repeated until the last agent has been selected. Once all agents have been queried, the principal uses the final report (from the last agent) and a decision rule to select an action. When the selected action is executed and the outcome is observed, the scores for all agents can be calculated, and the time step concludes. 
 
Note that while the principal faces a contextual Bernoulli bandit problem, every other agent faces a continuous contextual bandit problem, where the agent’s action is its probabilistic report to the principal (see Figure \ref{fig:three_diagrams}). To clearly distinguish between these two contextual bandit problems and the, we refer to the context of the Bernoulli bandit problem as the system’s context, and the context in the continuous bandit problem of the individual agents as the agent’s context. The agent’s context consists of the signals it receives from the system’s environment, and the previous report it receives from the principal or the previous agent. The system’s context consists of all signals that are received by the agents from the environment, including the priors that the principal receives from the environment.  The principal in this system cannot learn. However, the agents can learn to use the context to generate reports that maximise the score they receive. We test if, in such a system, the agents can efficiently learn such that the principal’s performance in the Bernoulli bandit problem improves.

In the following, we provide the notation and properties of the Bernoulli bandit problem, the agents and the context they receive, the principal’s  decision rule and scoring rule, and the agents' learning algorithm.

\subsection{Bernoulli bandit problem} \label{sec:bernoulli_bandit_problem}
We denote the time step as $T\in\{1,2,\ldots,n\}$. We assume that in each time step the principal selects one action from a finite, discrete set of action $A\in\{1,\ 2,\ldots,k\}$. The outcome $\Omega_{\left(A\right)}\ \in\{0,1\}$ of action $A$ is a Bernoulli variable. We assume $\Omega_{\left(A\right)}=1$ is the outcome desired by the principal. Selecting one action and observing the outcome will not reveal any information about the outcomes of the other actions.

We consider $m$ agents, which are denoted as $E\in\{1,2,\ldots,m\}$. For each time step $T$, agent $E$ will privately receive a signal, which provides information about the outcome of the principal’s available actions. The agent also receives a probabilistic report $Pr_{\left(E-1\right)\ }\in\left[0,1\right]^k$ from the principal. The agent’s context denoted by $C_{\left(E\right)}$ consists of the signals and the report from the principal and is used by the agent to return a report $Pr_{\left(E\right)}$. The probabilistic report about any specific action A is denoted as $Pr_{\left(E,A\right)}$.

\subsection{Principal, decision rule and scoring rule}
At the beginning of each time step $T$, the principal starts with an initial vector of probabilities $Pr_{\left(0\right)}$. This can be seen as prior probabilities that are provided from the environment to the principal, which is compatible with a common prior assumption that is often used in related work \cite{Pennock2007ComputationalMarkets}. The principal passes this initial report to the first agent, and the first agent returns an updated report denoted as $Pr_{\left(1\right)}$. The principal passes the updated vector to the next agent and repeats this procedure until the final report $Pr_{\left(m\right)}$ is received from the last agent $m$. After enquiring with all the agents, the principal uses the final report and a decision rule to select an action. We define a decision rule as a function which maps the final reports to a probability distribution over the available actions:
\begin{equation}
    \Phi:Pr_{\left(m\right)}\rightarrow\Delta\left(\{A\}\right)
\end{equation}
We denote $\Phi_A\left(Pr_{\left(m\right)}\right)$ as the probability of action $A$ to be selected. The principal will sample an action from the distribution $A \sim \Phi\left(Pr_{\left(m\right)}\right)$ and execute it. Afterwards, the principal observes the outcome $\Omega_{\left(A\right)}$ for the executed action and receives the corresponding reward. The principal will score the report of agent $E$ using a decision score function:
\begin{equation}
    S:Pr_{\left(E\right)}\times\Phi\left(Pr_{\left(m\right)}\right)\times A\times\Omega_{\left(A\right)}\rightarrow\mathbb{R}
\end{equation}
to compute the score for the report $Pr_{\left(E\right)}$ from agent $E$ given that the probabilities used to sample the action are $\Phi\left(Pr_{\left(m\right)}\right)$, the selected action is $A$ and the outcome is $\Omega_{\left(A\right)}$. For simplicity, we will omit the last three inputs and denote the decision scoring rule function as $S\left(Pr_{\left(E\right)}\right)$. 

The relationship between a decision scoring rule and a scoring rule is:
\begin{equation}
    S\left(Pr_{\left(E\right)},\Phi\left(Pr_{\left(m\right)}\right),A,\Omega_{\left(A\right)}\right)=\frac{1}{\Phi_A\left(Pr_{\left(m\right)}\right)}\hat{S}\left(Pr_{\left(E\right)},\Omega_{\left(A\right)}\right)
\end{equation}
where $\hat{S}\left(Pr_{\left(E\right)},\Omega_{\left(A\right)}\right)$ is any strictly proper scoring rule, such as the logarithmic scoring rule or the Brier score, and the decision rule $\Phi\left(Pr_{\left(m\right)}\right)$ has full support. In other words, strictly proper decision scoring rules calculate a proper score for a certain action and outcome by scaling up the inverse of the action’s probability from the decision rule. As a result, the expected scores from strictly proper decision rules do not depend on the actions’ probabilities from the decision rule. We further define $S\left(Pr_{\left(E\right)},\Phi\left(Pr_{\left(m\right)}\right),A,\Omega_{\left(A\right)}\right)=\ 0$ when $\Phi_A\left(Pr_{\left(m\right)}\right)\ =\ 0$ to include decision rules without full support, such as deterministic decision rules which we will discuss in Section 4.

\subsection{Continuum-armed contextual bandit learning}
As outlined in Section \ref{sec:bernoulli_bandit_problem}, each agent receives a report from the previous agent (or the principal if the agent is the first agent to report) and a private signal from the environment. Based on this context the agents return an updated report which is scored. The updated report represents an agent’s action, and therefore we can treat each agent as a continuum-armed contextual bandit agent. In many learning algorithms, the continuum-armed bandit problem is discretised to a finite-armed bandit problem \cite{Agrawal1995TheProblem, Kleinberg2005NearlyProblem}. In this study, we treat this problem differently, by learning parameters which generate updated reports with the policy gradient method that maximises the expected score of agent $E$,  $\mathbb{E}\left[S\left(Pr_{\left(E\right)}\right)\right]$ \cite{Williams1992SimpleLearning}.

Formally, at time step $T$, we assume that agent $E$ keeps a matrix of parameters $\Theta_{\left(E\right)}$. Given the context $C_{\left(E\right)}$ and the prior probability $Pr_{\left(E-1\right)}$, the agent will construct a $k$ dimensional density function. We assume that the density function is a k-dimensional normal distribution $\pi=\ N\left(\mu\left(C_{\left(E\right)},Pr_{\left(E-1\right)},\Theta_{\left(E\right)}\right),\sigma^2\right)$ with the means $\mu\left(C_{\left(E\right)},Pr_{\left(E-1\right)},\Theta_{\left(E\right)}\right)$. After sampling a vector of log-odds $H_{\left(E\right)} \sim N\left(\mu\left(C_{\left(E\right)},Pr_{\left(E-1\right)},\Theta_{\left(E\right)}\right),\sigma^2\right)$ from this probability density, the agent computes the updated probabilistic report as:
\begin{equation}
    Pr_{\left(E,A\right)}=\frac{1}{1+\exp{\left(-H_{\left(E,A\right)}\right)}}
\end{equation}
for action $A$.

At time step $T$ agent $E$ updates its parameters to maximise the expected score by gradient ascent
\begin{equation} \label{eq:gradient_update}
\Theta_{\left(E,T+1\right)}=\Theta_{\left(E,T\right)}+\alpha\frac{\partial\mathbb{E}\left[S\left(Pr_{\left(E,T\right)}\right)\right]}{\partial\Theta_{\left(E,T\right)}}
\end{equation}
where $\alpha$ is the learning rate. The approximation of the gradient from an agent’s experience follows the methodology described in \cite{Williams1992SimpleLearning, Sutton2018ReinforcementIntroduction}. Further detail is provided in Section \ref{sec:simulation_setup}.

\subsection{Simulation setup} \label{sec:simulation_setup}
In the simulations, we use the classic urn problem as the model for the Bernoulli bandit problem. Specifically, the principal will face an environment which consists of k urns. Each urn represents an action of the k actions in the Bernoulli bandit problem. There are two types of urns, which are red type (1) and blue type (0), representing the possible outcomes of the action, i.e. the Bernoulli variable. The type is hidden from the principal until the principal selects the urn. The type of the other (unselected) urns, however, remains hidden. 

At the beginning of each time step $T$, $k$ prior probabilities $Pr_{\left(0\right)}$ will be sampled from a normal distribution (in log-odds form), one for each urn. The Bernoulli type of urn $A$ will be sampled using the prior probability $Pr_{\left(0,A\ \right)}$. The prior probabilities will be given to the principal and they will be re-sampled at each time step. Urns contain multiple balls of two colours, red and blue. The composition is determined by the type of the urn and remains fixed for all time steps $T$. Simulations in section \ref{sec: simulation results} are all conducted in an environment with two urns each of which can be Bernoulli type 0 or 1. A Bernoulli type 1 urn contains 2/3 red balls and Bernoulli type 0 contains 1/3 red balls.

A number of $J$ balls will be randomly sampled with replacement from one or multiple urns by an agent. The colour of these balls constitutes the private signal of the agent from the environment. The colour of the balls (and their origin), as well as the prior probabilities $Pr_{\left(0\right)}$ or the previous updated report $Pr_{\left(E-1\right)}$ jointly form the contextual information vector $C_{\left(E\right)}$ of an agent. For our setting with two urns to select from, two types of urns and balls, the context can be implemented as a vector with 6 elements (see equation \ref{eq:matrix}), where $c_{r1}$ represents the number of red balls drawn from urn 1. $c_{b1}$ represents the number of blue balls drawn from urn 2, and $c_{p1}$ is the log odds transformed prior report for urn 1. Similarly, $c_{r2}$, $c_{b2}$ and $c_{p2}$ represent the number of red and blue balls as well as the log odds transformed prior report for urn 2.

The contextual information vector multiplied with the matrix of learning parameters $\Theta_{\left(E\right)}$ of the agent gives the means $\mu\left(C_{\left(E\right)},\Theta_{\left(E\right)}\right)$ for a normal distribution $N \sim \left(\mu\left(C_{\left(E\right)},\Theta_{\left(E\right)}\right),\sigma^2\right)$. The log odds $H_{\left(E\right)}$ of the actual report will be sampled from this normal distribution. The computation of the means for the updated report can be written as

\begin{equation} \label{eq:matrix}
\begin{pmatrix}
c_{r1} \\
c_{b1} \\
c_{p1} \\
c_{r2} \\
c_{b2} \\
c_{p2} 
\end{pmatrix}^\intercal  \times
\begingroup
\renewcommand*{\arraystretch}{1.5}
\begin{pmatrix}
\theta^{(1)}_{r1} & \theta^{(2)}_{r1} \\
\theta^{(1)}_{b1} & \theta^{(2)}_{b1} \\
\theta^{(1)}_{p1} & \theta^{(2)}_{p1} \\
\theta^{(1)}_{r2} & \theta^{(2)}_{r2} \\
\theta^{(1)}_{b2} & \theta^{(2)}_{b1} \\
\theta^{(1)}_{p2} & \theta^{(2)}_{p2} 
\end{pmatrix}
\endgroup
 =
\begin{pmatrix}
\mu_1 & \mu_2 
\end{pmatrix}  
\end{equation}

The result of the multiplication $\mu_1$ is the mean log-odds transformed report for urn 1, and $\mu_2$ is the mean log-odds transformed report for urn 2. The reported log odds for urn 1 and urn 2 will be sampled from a normal distribution with this mean and a fixed variance, i.e., $H_1 \sim N\left(\mu_1,\sigma^2\right)$, and $H_2 \sim N\left(\mu_2,\sigma^2\right)$.

We implement logarithmic scoring rules in the simulation. Therefore, the decision scoring rule can be written as

\begin{equation}
\begin{split}
S(Pr_{\left(E\right)},& \Phi\left(Pr_{\left(m\right)}\right),A,\Omega_{\left(A\right)}) \\ &= 
\begin{cases}
     \frac{1}{\Phi_A\left(Pr_{\left(m\right)}\right)}\log{\frac{Pr_{\left(E\right)}}{Pr_{\left(E-1\right)}}}, &\text{if $\Phi_A\left(Pr_{\left(m\right)}\right) > 0$ and $\Omega_{(A)} = 1$} \\
     \frac{1}{\Phi_A\left(Pr_{\left(m\right)}\right)}\log\frac{1-Pr_{\left(E\right)}}{1-Pr_{\left(E-1\right)}}, &\text{if $\Phi_A\left(Pr_{\left(m\right)}\right) > 0$ and $\Omega_{(A)} = 0$} \\
    0, &\text{otherwise} \\
\end{cases}    
\end{split}
\end{equation}

This implies that an agent receives a difference in the logarithmic score between the agent’s score and the previous report’s score. In other words, the agent is scored for how much the accuracy of the preceding report is improved or worsened. In our simulation, we use a stochastic decision rule that favours the type one urn. In other words, we assign the highest probability to the urn that is forecasted to be most likely to be type one. In our two urns simulations, we assign a probability of 90\% to selecting probability to the urn that is reported to be most likely type one and 10\% to the other. From the principal perspective, this is essentially an $\epsilon$ greedy two-arm bandit problem with $\epsilon=\ 10\%$. A fixed exploration rate will cause a loss of performance, which is clear in our simulation results in section \ref{sec:decision_systems}. One can use bandit learning techniques to optimise the principal, but this is not the focus of this work.

Agent E’s initial parameters $\Theta_{\left(E\right)}$ is sampled from a standard normal distribution $N \sim \left(0,1\right)$. At time step T, we refer a tuple $\left(C_{\left(E,T\right)},\mu_{\left(E,T\right)},H_{\left(E,T\right)},S\left(Pr_{\left(E,T\right)}\right)\right)$ that consists of useful information for agent $E$ to learn as an experience. We use the experience replay buffer technique as in \cite{Lin1993ReinforcementNetworks, Mnih2013PlayingLearning}. The difference is, in our simulation, each experience is independently and identically distributed, and therefore we apply this technique only for the efficiency of the hardware usage. After agent $E$ receives the score, it will store the experience in its own experience replay buffer. If the existing experience number exceeds a certain threshold, the latest experience will replace of the oldest one. Afterwards, a fixed number $F$ of experience tuples will be uniformly sampled from the experience replay buffer for an update. Assume the experience tuple at time step $I$ is within that $F$ samples, the gradient for the tuple can be obtained by
\begin{equation}
    G_{\left(E,I\right)}=C_{\left(E,I\right)}\times\left(S\left(Pr_{\left(E,I\right)}\right)-B\left(C_{\left(E,I\right)}\right)\right)\times\frac{H_{\left(E,I\right)}-\mu_{\left(E,I\right)}}{\sigma^2}
\end{equation}
where $B(C_{(E,I)})$ is a baseline function that does not vary with action $A$ but only depends on the contextual vectors \cite{Sutton2018ReinforcementIntroduction}. Finally, an average gradient of a mini-batch will be computed by 
\begin{equation}
    \overline{G_{\left(E,T\right)}}=\frac{1}{F}\sum_{I} G_{\left(E,I\right)}
\end{equation}
to update the parameters of time step T as shown in equation \ref{eq:gradient_update}.

\subsection{Performance evaluation}
The principal’s objective is to select urns of type 1. To track the systems' performance we, therefore, use the sum of  Bernoulli outcome variable $\sum_{T=1}^{n}\Omega_{\left(A,T\right)}$ of the executed action A. Additionally, we use the error of the final reports, defined as the mean squared residual between the final aggregated reports that the principal receives from the sequential reporting and the correctly updated Bayesian posterior $\widehat{Pr}$ of an observer with access to the entire environmental context.
\begin{equation}
    Er=\sum_{A}\left(Pr_{\left(m,A\right)}-\widehat{Pr}\right)^2
\end{equation}
This metric is used only for evaluation purposes as the signals are not accessible by the principal during the training phase.

\section{Simulation results} \label{sec: simulation results}
We use our simulation setup to investigate three different decision market scenarios. In the first set of simulations (Section \ref{sec:decision_systems}) we compare a multi-agent system with a centralised agent. In the multi-agent system, signals are distributed across the individual agents, while in the centralised system, there is a single agent that receives all signals. In both cases, a stochastic decision rule is used. The results show that the multi-agent contextual bandit system performs as well as the centralised system. 

In the second set of simulations (Section \ref{sec:decision_simulation}), we analyse decision markets with deterministic decision rules, starting with a single agent. The results show that because such markets are not incentive-compatible, agents can learn strategies that lead to reports which differ from correct information aggregation. The agent’s behaviour resembles strategies described by \citeauthor{Othman2010DecisionMarkets} \cite{Othman2010DecisionMarkets}. However, in many cases, agents learn to provide accurate forecasts (despite receiving a lower reward), which indicates that strategically inaccurate forecasting is difficult to learn with gradient methods.

In the third set of simulations (Section \ref{sec:decision_agents}), we are investigating decision markets with deterministic decision rules and multiple agents. We observe novel strategies leading to non-trivial interactions between agents, with strategically distorted reporting by the agent who reports first. The results show that while final reports are as accurate as for decision markets with stochastic decision rules, the distribution of rewards for agents under stochastic decision rules is fairer compared to the distribution under deterministic decision rules. 
\subsection{Decision markets with stochastic decision rules: distributed vs. centralised systems } \label{sec:decision_systems}

In this set of simulations, we compare the performance of a system with $J$ individual agents, each of which receives a single signal, with a corresponding centralised system where a single agent receives $J$ signals. The simulations follow the approach described in Section \ref{sec: algorithm}, with $J$ being set to 3, 5, 9, and 15. Each signal is a draw of a single ball (sampled with replacement) from one of the urns.

\begin{figure}
    \centering
    \includegraphics[width=\textwidth]{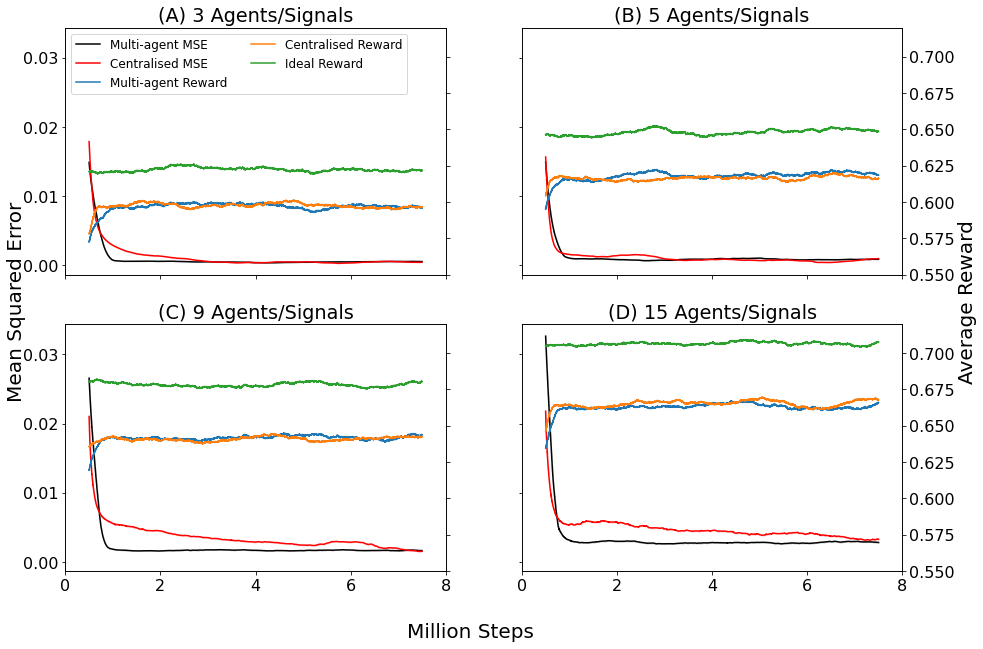}
    \caption{System performance of multi-agent and centralised systems. Panel (A)-(D) show simulations with 3,5,9 and 15 signals. In the multi-agent system, each agent receives one signal. In the centralised counterpart, a single agent receives all signals. The black line and red line are the running averages of the mean squared error of multi-agent and centralised systems, respectively. The green line is the average received reward for a principal who chooses an action according to a posterior from a correct Bayesian model that can use all available information. The blue line shows the actual reward received for the multi-agent systems. The orange line is the reward received by the centralised systems. The errors and rewards for the centralised and distributed systems are very similar. The rewards are lower compared to the Bayesian model, with the difference arising from the use of a stochastic decision rule in the multi-agent and centralised system.}
    \label{fig:distributed_centralised}
\end{figure}

As shown in Figure \ref{fig:distributed_centralised}, we observe that the mean square error (MSE) of the final report decreases rapidly and stabilises close to zero in both multi-agent and centralised systems. The MSE declines faster in the multi-agent system, compared to the centralised counterpart when the agent or signal number is high. This is analysed in more detail further below. Once converged, the average rewards for both systems are very similar, with the reward being defined as one when the selected urn turns out to be of the preferred type (red), and zero when it is not. Note that the gap between the actual reward and the ideal reward is due to the nature of stochastic decision rules, which assigns a positive probability to select a sub-optimal action. The performance will be close to the ideal reward if we account for the disadvantage of the stochastic decision rules. 

\begin{figure}
    \centering
    \includegraphics[width=\textwidth]{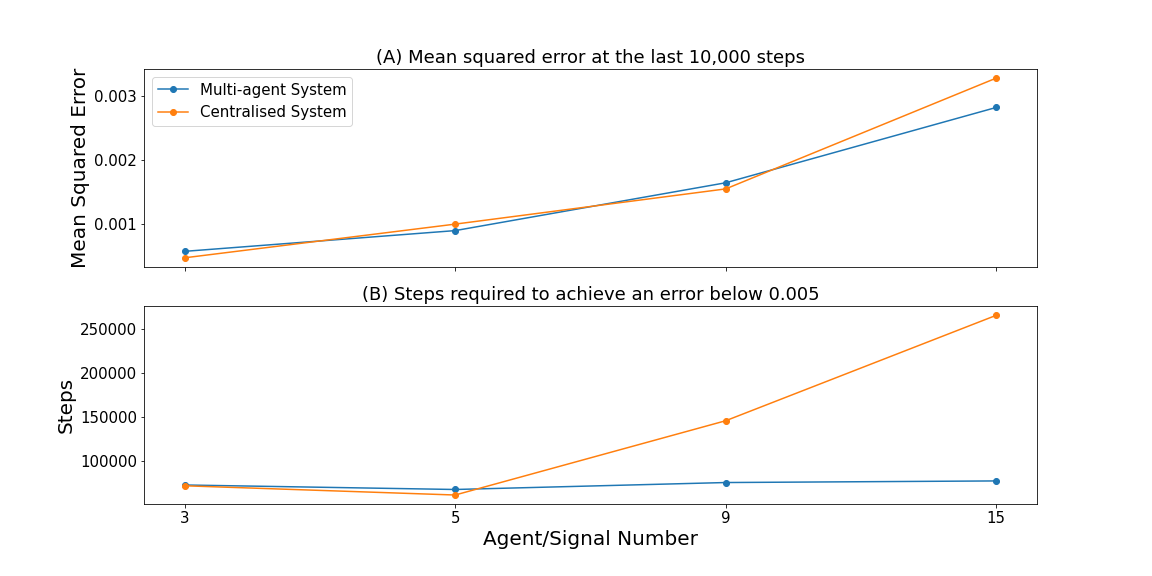}
    \caption{Final accuracy and time to convergence in multi-agent and centralised systems. Plot (A) shows the mean squared error in the last 10,000 steps. Plot (B) shows the training iterations required to reach a mean squared error below 0.005.}
    \label{fig:converge_speed}
\end{figure}

We further analyse performance by investigating the time to convergence, and the average squared residual error after convergence. We recorded the average MSE at the last 10,000 steps (see Figure \ref{fig:converge_speed}A) and the number of training steps (see Figure \ref{fig:converge_speed}B) required for a system to reach an acceptable performance (here set to 0.005). Figure \ref{fig:converge_speed}A shows the converged performance of both systems increases similarly with an increasing number of signals. Figure \ref{fig:converge_speed}B, however, indicates that the steps required for the centralised system to reach an acceptable performance increase with the number of signals it receives. In contrast, the multi-agent system does not show any relation between the number of agents and the steps required for training to reach a MSE of 0.005. In other words, the multi-agent system shows better scalability.

\begin{figure}
    \centering
    \includegraphics[width=\textwidth]{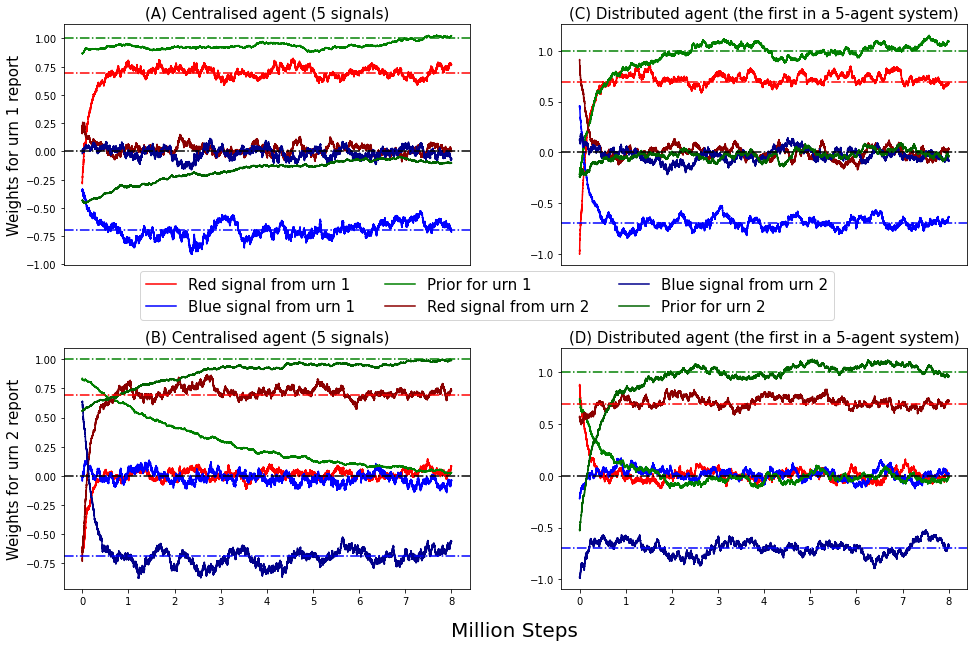}
    \caption{Progress of the learning parameters for a centralised and distributed agent. Plot (A) and plot (B) show the six parameters that determine the report for urn 1 and urn 2 respectively for a centralised agent. Plot (C) and plot (D) show the parameters for a distributed agent. These parameters are the left column in the matrix in equation \ref{eq:matrix}. Specifically, the red line with the label “Red signal from urn 1” in the legend is the history of $\theta_{r1}^{\left(1\right)}$ and the parameter can be interpreted as the weight of receiving a red ball from urn 1 for the posterior report of the urn 1. Similarly, “Blue signal from urn 1”, “Red signal from urn 2” and “Blue signal from urn 2” show the weights corresponding to the signal colour and which urn it comes from. The green lines labelled “Prior for urn 1” and dark green lines labelled “Prior for urn 2” show the weights for prior probabilities passed by the previous agent or the principal on the updated reports. The dash lines are the ideal value for the learning parameters.}
    \label{fig:distributed_centralised_weights}
\end{figure}

Figure \ref{fig:distributed_centralised_weights} shows the progress of learning parameters $\Theta$ for a centralised agent (panels A and B) and a distributed agent from a multi-agent system (panels C and D). As shown in the figure, the learning parameters converge to theoretically ideal values. For a centralised agent, the parameters that relate prior probabilities with reports converge slower than in the distributed counterpart. This is because compared to the contribution of five balls to the final report, the information provided by the prior probability is less important. The time to convergence increases when the signal number increases, while in the multi-agent counterpart each agent learns every parameter at a similar pace.

Overall, from the simulation results, we find that a multi-agent bandit system that uses a decision market with stochastic decision rule can learn to make highly accurate reports, similar to the corresponding centralised system. The advantage of our multi-agent system is that the contextual information an agent receives can remain private to the agent. More specifically, the agents reveal how the contextual information they receive affects their probabilistic reports, but they do not need to reveal the contextual information itself, or the weights that were used to link the context with the report.

\subsection{Decision markets with deterministic decision rules: single agent simulations} \label{sec:decision_simulation}

Section \ref{sec:decision_systems} demonstrates that decision markets with stochastic decision rules can elicit information distributed over multiple agents. These agents can be computational and can use the decision market score to learn using their information to make accurate forecasts. However, decision markets with stochastic decision rules are inefficient because they entail that the principal sometimes selects an action that is forecasted not to be the best possible action. It is in the interest of the principal to use a deterministic decision rule and select the action that has the highest probability to achieve the desired outcome. 

Such a decision rule, however, is not incentive-compatible and has been theoretically shown to be manipulatable by rational and myopic agents \cite{Othman2010DecisionMarkets, Chen2011InformationMaking}. From a reinforcement learning perspective, a score derived from a decision market with a deterministic decision rule cannot be expected to allow agents to learn providing reports that can be interpreted as accurate probabilistic reports. \citeauthor{Othman2010DecisionMarkets}’s work discusses strategies of an agent who is the last to make a report and can benefit from strategically inaccurate reporting, and show that there are situations where subsequent agents with the same piece of information have no incentives to correct such inaccuracies. 

We here investigate the strategies that are learned by a single agent with a single signal as specified in Section \ref{sec: algorithm}. Further simulations with multiple agents with independent information are investigated in Section \ref{sec:decision_agents}. 

\begin{figure}[ht]
    \centering
    \includegraphics[width=\textwidth]{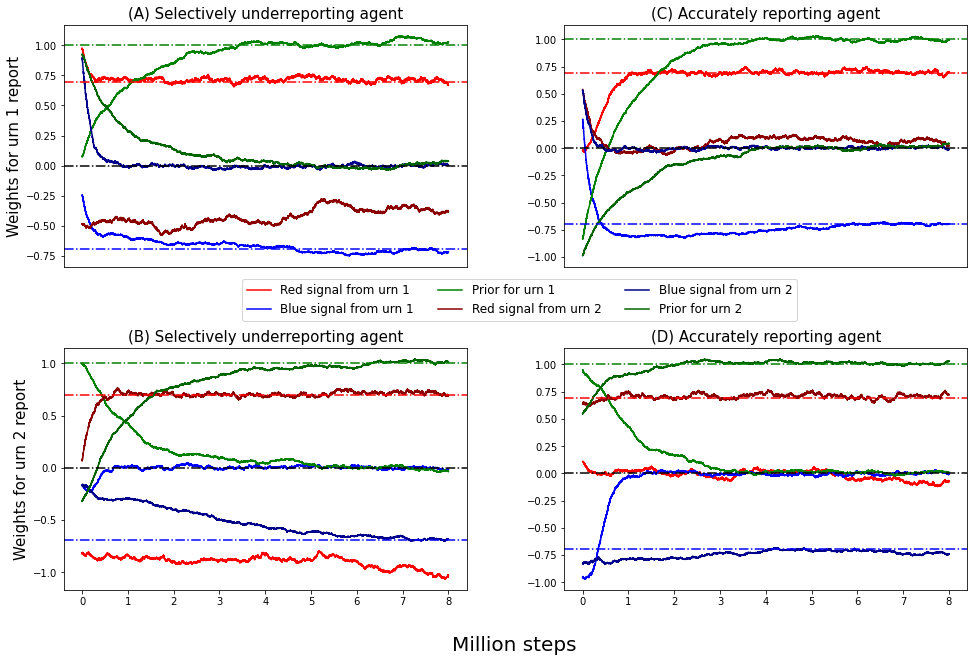}
    \caption{Progress of the learning parameters for a selectively underreporting and accurately reporting agent. Plot (A) and (B) are weights for the log odds report from a selectively under-reporting agent. Plot (C) and (D) are weights from an accurate reporting agent. The accuracy is determined by if the posterior report agrees with the posterior Bayesian inference with all the signals accessible.}
    \label{fig:lying_honest_weights}
\end{figure}

We find that in our simulations of a single agent in a decision market with a deterministic decision rule, depending on initial parameters different strategies are learned. In most simulations, agents learn to make reports similar to those in the stochastic decision markets. The reports of the agents represent accurately estimated probabilities for the outcomes, given the information available. Such agents could be seen as ‘honest‘ agents. We also observe agents learn accurately report the probability for one urn, but provide a report that is lower than the accurate report for the other urn. The weights of an accurately reporting agent and a ‘selectively underreporting’ agent are shown in Figure \ref{fig:lying_honest_weights}.

While the weights for an accurate agent are the same as the ones learned by the agents in the decision markets with stochastic decision rule, the selectively underreporting agent differs in two weights. These two weights are zero for the accurate agents but negative for the selectively underreporting agent. One of these weights lowers the report for urn 2 when a red signal for urn 1 is received; the other weight lowers the report for urn 1 when a red signal for urn 2 is received. This contrasts with an accurate agent who would not lower a report for an urn for which no signal has been received.

\begin{figure}[ht]
    \centering
    \includegraphics[width=\textwidth]{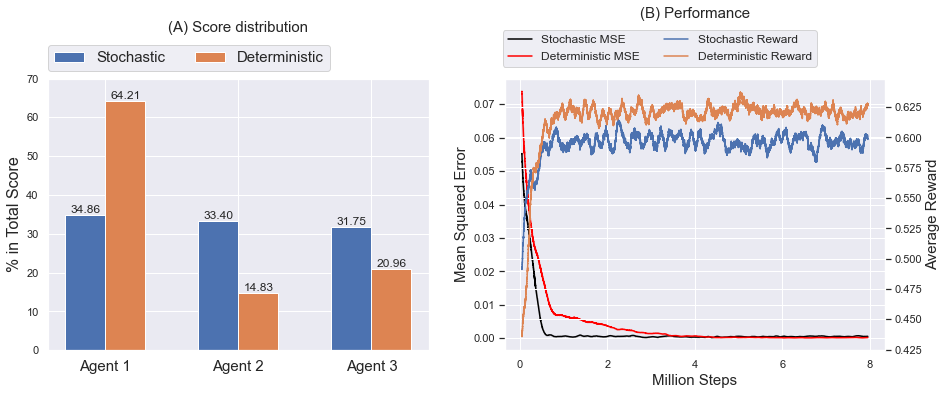}
    \caption{Reward distribution and performance of decision markets with stochastic and deterministic decision rules. Plot (A) shows the average score ratio comparison between individual agents in different reporting sequences and different decision markets. Plot (B) shows the performance comparison between a three-agent system based on a decision market with a stochastic and deterministic decision rule.}
    \label{fig:single_agent_performance}
\end{figure}

A selectively underreporting agent can benefit from this strategy (see Figure \ref{fig:single_agent_performance}) because if reporting accurately, the agent receives a larger payoff if the urn is selected from which the signal was received. A single agent can therefore maximise its payoff by reporting accurately for the urn from which the signal was received, and submitting a report for the other urn that is sufficiently low such that the former rather than the latter urn is selected. However, we only observe agents learn to selectively underreport when obtaining a red signal. In principle, selective underreporting also maximises the payoff for an agent who receives a blue signal. However, the blue signal lowers the probability for an urn to be of the favourable type; therefore this strategy requires lowering the report for the other urn much more. If underreporting is insufficient to change the choice of the urn, it is disadvantageous, making selective underreporting difficult to learn with local, gradient-based methods when a blue signal is received.

In summary, in our single-agent simulations, we find strategic manipulation similar to the strategies expected by \citeauthor{Othman2010DecisionMarkets}. An agent can benefit from manipulating the final report and misleading the principal to a sub-optimal action. However, such a strategy would be vulnerable to subsequent agents with the same piece of the information who have incentives to correct previous underreporting. 

\subsection{Decision markets with deterministic decision rules: simulations with multiple agents} \label{sec:decision_agents}

The multi-agent dynamic in a deterministic decision market has so far not been investigated. We here use our system to investigate the interaction between three agents in a  decision market with a MAX decision rule.

In Section \ref{sec:decision_simulation}, we observe agents learn to ‘game’ a decision market by ‘honestly’ reporting about the urn that offers the larger expected reward to the agents while underreporting (or ‘trash-talking’) the other option. This strategy maximises expected payoffs but could be exploited by subsequent agents. \citeauthor{Othman2010DecisionMarkets} discuss a strategy where the final agent strategically inflates the final report of the urn that offers a higher reward than accurately reporting to the agent under certain conditions (see example 2 in \cite{Othman2010DecisionMarkets}). Such a strategy can also be profitable, and it cannot be exploited by subsequent agents with the same information. However, it is unclear what strategies are beneficial when multiple agents with conditionally independent signals exist. To study this situation, we use simulations with three agents. 
As in the previous simulations (Section \ref{sec:decision_systems}), each agent will draw a ball from a random urn and return it after privately recording the colour. The agents make sequential reports. 

\begin{figure}[ht]
    \centering
    \includegraphics[width=\textwidth]{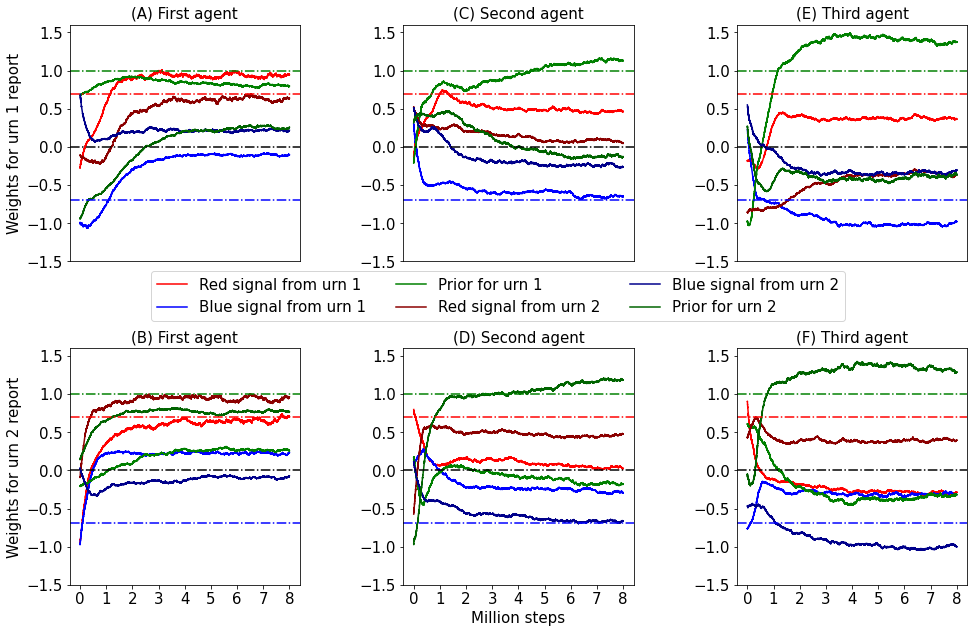}
    \caption{Progress of the learning parameters for a three-agent system based on a decision market with a deterministic decision rule. The $2 \times 3$ grid plots show the change of weights from each agent. The dashed lines show the weights required for strict Bayesian updates like reports. The weights in panels (A) and (B) show the weights of the first agent in a decision market with a deterministic decision rule; (C) and (D) of the second agent; and (E) and (F) of  the final agent.}
    \label{fig:three_agent_weights}
\end{figure}

\begin{figure}[ht]
    \centering
    \includegraphics[width=\textwidth]{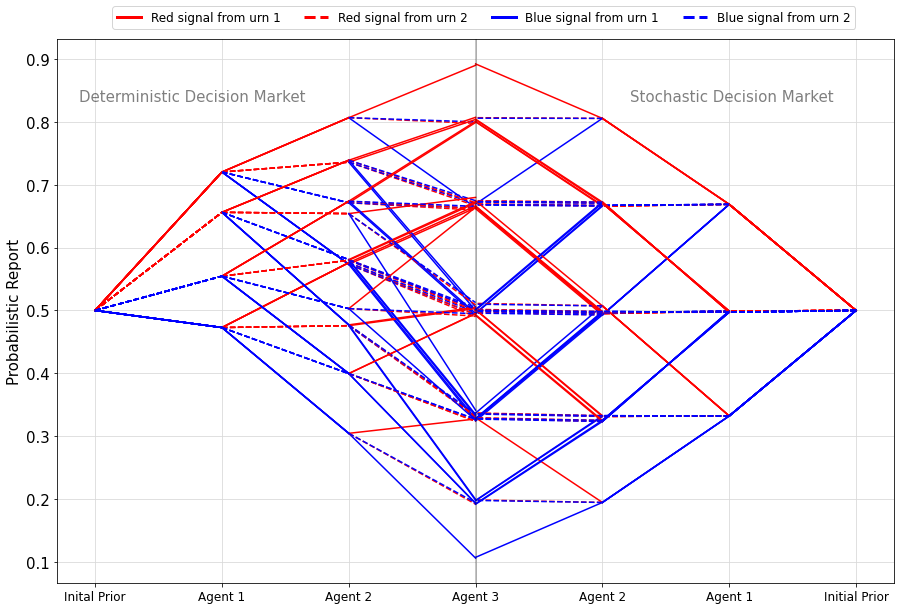}
    \caption{Change of probabilistic reports in decision markets with deterministic and stochastic decision rules. The plot has shown the change of the probabilistic report for urn 1 passed by each agent (the report for urn 2 is symmetric). A line labelled “Red signal from urn 1” shows the change in the report after receiving a red ball from urn 1. The other labels follow the same convention. A solid line indicates the ball is from urn 1 (the reporting urn) and a dashed line means the ball is from urn 2 (the other urn).}
    \label{fig:diamond}
\end{figure}

Figure \ref{fig:three_agent_weights} shows the agents' strategies in terms of the weights learned by the agent. The weights for the first agent are all increased compared to a Bayesian agent accurately reporting the probabilities given the information. This means that the first agent strategically overreports. Subsequent agents essentially correct this initial overreporting such that the final reports become quite accurate (see Figure \ref{fig:diamond}.). Intuitively, this strategy is beneficial because the first agent learns that the urn with the highest probability will be selected. By providing increased reports on all outcomes, the first agent can benefit from increasing all reports. The agent increases the report most when receiving a red signal, but it also increases the reports for the urn from which no signal has been received. When a blue signal is received, it provides the lowest report, though this report is still larger than one given by an accurate Bayesian agent (see Figure \ref{fig:diamond}.). Interestingly, this leads to a reward distribution that substantially favours the first agent. While agents in the decision markets with stochastic decision rules receive very similar rewards, under a deterministic decision rule, the first agent receives a much higher expected score. Thus, while from the principal’s perspective the decision making performance is very similar to the strategies emerging in our simulations, the expected scores offered under a stochastic decision rule can be seen as more ‘fair’ compared to the scores under a deterministic decision rule.

With the simulation, we reveal interesting dynamics in a multi-agent system with a deterministic decision market. The first agent learns an overreporting strategy and takes the lion's share of the score. The subsequent agents correct the report of the first agent, which results in a surprisingly accurate final report.

\section{Conclusion and discussion} \label{sec: conclusion}

In this paper, we investigate the use of decision markets, which are economic mechanisms for decision-making based on distributed information, for contextual bandit learning in a multi-agent system. Unlike existing multi-agent systems, we assume contextual information is distributed across multiple self-interested agents who own their information and require incentives to reveal it and learn to interpret it. This scenario is relevant for real-world commercial models because contextual information such as user profiles or patient information is often proprietary and potentially too sensitive to be made available to a centralised model. Rather than revealing contextual information, training data and learned parameters, each agent solely need to reveal its predictions for the available actions. Because a decision market offers a proper score for these predictions, it allows training multiple self-interested models without accessing private contextual information. The score aligns the individual agents with the system's performance such that as the agents improve their individual scores, the collective decision making efficiency improves as well.

Our simulations show that the decision market based multi-agent system can train self-interested agents to achieve an equally efficient performance as a centralised trained counterpart with accessibility to all pieces of the same contextual information. This result indicates that coordination problems and the non-stationary issue that can arise in multiagent systems do not affect the performance of our system in the simulations.

We use our system to investigate the dynamics of multi-agent interactions under different decision rules. While decision markets with stochastic decision rules allow the agents to learn to make highly accurate forecasts, the stochastic decision rule reduces the efficiency of decision making. We, therefore, simulate how agents learn under a deterministic decision rule. In the one-agent system with a deterministic decision rule, the agent, who learns with a gradient-based algorithm, can learn to selectively diminish the probability of the action for which it does not have any information. Moreover, the agent benefits from this underreporting strategy because it ensures that the action is selected for which the forecasts are expected to be scored higher. However, the learned strategies depend on the initial values for the learning parameters, and often the agents learn to report accurately. This highlights the limitations of gradient methods used in the simulations to find the global optimum.

In a three-agent system with a deterministic decision rule, we observe the first agent learn to overreport for both actions and thereby gain a significant first-mover advantage. The subsequent agents gradually correct the report which results in an accurate final report. The average scores for individual agents are less equitably distributed under a deterministic decision rule, compared to a stochastic decision rule. Our results suggest that our simulation-based approach to testing economic mechanisms in a multiagent learning context can identify strategies that are beneficial to the individual agents and their consequences for the overall system performance.

A future study could use global optimising techniques to find globally optimal strategies and thereby help identify Nash equilibria. For instance, in Section \ref{sec:decision_simulation} we mentioned that selective underreporting for a blue signal is difficult to learn with local gradient-based methods. This is because underreporting has to be sufficient to change the decision. Less strong underreporting does not change the decision but reduces accuracy for the selected action and therefore reduces the score. Global optimising techniques, however, have a much higher computational complexity. Another future study direction is overcoming the limitation of stochastic decision rules, which sometimes require the principal to select the action that is predicted to be sub-optimal. This requires a mechanism that allows for a deterministic action selection while simultaneously maintaining incentive compatibility. A promising approach might be to use peer prediction methods to resolve the decision markets.

\bibliography{multi-agent_system,bandit_rl,federated_learning,decision_markets}
\bibliographystyle{unsrtnat}

\end{document}